\documentclass[a4paper,twocolumn,twoside,showpacs,showkeys, preprintnumbers,superscriptaddress,amsmath,amssymb,pra]{revtex4}
\usepackage[dvips]{graphicx}
\usepackage{xspace}
\usepackage[dvips]{color}

\definecolor{grey}{gray}{0.8}

\newcommand{\comment}[1]{}

\newcommand{\ket}[1]{\ensuremath{|#1\rangle}}

\newcommand{\ketbra}[3][]{\ensuremath{|#2\rangle_{\!#1\!}\langle#3|}}

\newcommand{\gone}{\ensuremath{g_{(1)}}}
\newcommand{\gtwo}{\ensuremath{g_{(2)}}}
\newcommand{\capitallabel}[1]{\textsc{\small\mbox{#1}}\xspace}
\newcommand{\spdc}{\capitallabel{SPDC}}
\newcommand{\ktp}{\capitallabel{KTP}}

\newcommand{\fwhm}{\capitallabel{FWHM}}

\newcommand{\etal}{{\small\em{et\,\,al.}}\xspace}
\newcommand{\average}[1]{\ensuremath{\langle #1 \rangle}}
\newcommand{\vacuum}{\ket{\emptyset}}
\newcommand{\cohtime}{\ensuremath{\tau_c}}
\newcommand{\pumptime}{\ensuremath{\tau_p}}
\newcommand{\Gone}{\ensuremath{G_{(1)}}}
\newcommand{\Gtwo}{\ensuremath{G_{(2)}}}

\begin{document}

\title{Photon bunching in parametric down-conversion
       with continuous wave excitation}

\newcommand{\Faculty}{Quantum Optics, Quantum Nanophysics and
  Quantum Information, Faculty of Physics, University of Vienna,
  Boltzmanngasse~5, 1090 Vienna, Austria}
\newcommand{\ARC}{Austrian Research Centers~GmbH~-~ARC,
  Donau-City-Str.~1, 1220 Vienna, Austria}
\newcommand{\firstauthor}{\thanks{These authors contributed equally to the work.}}

\author{Bibiane Blauensteiner} \firstauthor
\author{Isabelle Herbauts}     \firstauthor
\author{Stefano Bettelli}      \firstauthor
\affiliation{\Faculty}
\author{Andreas Poppe}
\affiliation{\ARC}
\author{Hannes H\"{u}bel}
\email[Corresponding author:~]{hannes.huebel@gmail.com}
\affiliation{\Faculty}

\preprint{arXiv:0810.4785v2}

\date{\today}
\begin{abstract}

The first direct measurement of photon bunching ($\gtwo$ correlation
function) in one output arm of a
spontaneous-parametric-down-conversion source operated with a
continuous pump laser in the single-photon regime is
demonstrated. The result is in agreement with the statistics of a
thermal field of the same coherence length, and shows the feasibility
of investigating photon statistics with compact cw-pumped
sources. Implications for entanglement-based quantum cryptography are
discussed.
\end{abstract}

\maketitle


\section{Introduction}

Light sources based on {\em spontaneous parametric down-conversion} (\spdc)
\cite{BEZ_book}, in which photons from a pump laser are sporadically converted into
pairs of photons at lower frequency in a non-linear crystal, have become an
essential component in the toolbox of quantum optics laboratories in the last
decade. These sources allow state-of-the-art experiments on entanglement and
quantum information, due to the ability to supply photons that are highly
correlated in time, energy and polarisation.

\spdc-based systems have been exploited as a basis for
entangled-photons schemes \cite{KMWZ95} as well as approximations of
single-photon sources in the so-called heralded-photon schemes
\cite{HongMandel85}. Such setups are however limited by the
possibility that additional pairs are created within the window
determined by the pump-pulse duration or by the electronic gate width
of the single photon detectors (for continuous sources). In
multi-photon interference experiments, additional pairs in general
affect the purity of the investigated state \cite{ORW99}. In quantum
cryptography \cite{SBPCDL08} instead, multi-photon pairs may
constitute a security threat, since their information-carrying degree
of freedom (typically polarisation) might be correlated~\cite{DusekBradler02}.

The topic of \spdc-pair emission and correlation, deeply related to
the statistical properties of the down-converted field, is therefore
not only interesting for fundamental quantum optics experiments, but
also concerns applied quantum information. Significant theoretical and
experimental investigations have already been conducted on this
subject, but a direct demonstration that the field of one arm of an
\spdc source is thermal is still missing for the special case of a
source in {\em single-photon regime} operated with a {\em continuous
  pump}. Such evidence is presented in this communication by measuring
the second-order correlation function with a Hanbury Brown-Twiss setup
\cite{HBT56} and comparing it to a prediction based both on the detector
jitter and a first-order correlation measurement with a Michelson
interferometer. This approach relies entirely on experimentally
observable quantities and no assumptions are needed.

The intrinsic difficulty of the experiment stems from the fact that
resolving times are much larger than the coherence time $\cohtime$
of the down-converted field, so that all relevant quantities are
averaged out and the statistics becomes similar to a
Poissonian one, with the true (thermal) statistics often overlooked.

The remaining part of the introduction presents a
review of the theoretical background
(Sec.\ref{sec:stats}) and existing
experimental evidence (Sec.\ref{sec:experiments}) concerning
multi-pair emission, as well as a discussion of the potential security
risk for quantum cryptography (Sec.\ref{sec:QKD}). The experimental
setup, data acquisition (Sec.\ref{sec:setup}), and quantitative
expectation model (Sec.\ref{sec:model}) are then described, followed
by a discussion of experimental findings (Sec.\ref{sec:conclusions}).

\subsection{Field statistics and theoretical description of \spdc}
\label{sec:stats}

A direct characterisation of the {\em temporal} statistical properties
of a generic quantum electromagnetic field is provided by the
(normalised) correlation functions, defined as
\cite[ch.6]{Loudon_book}
\begin{subequations}
  \begin{align}
    \label{eq:G1}
    \gone(t,t+\tau) &= \frac{\average{ \hat{E}^\dagger(t+\tau)
        \hat{E}(t) }}{\average{\hat{E}^\dagger(t) \hat{E}(t)}}=\frac{
      \Gone(t,t+\tau)}{\Gone(t,t)} \\
    \label{eq:G2}
    \gtwo(t,t+\tau) &=\frac{\langle \hat{E}^\dagger(t) \: \hat{E}^\dagger(t+\tau) \:
    \hat{E}(t+\tau) \: \hat{E}(t)\rangle} {\langle \hat{E}^\dagger(t)\:
    \hat{E}(t) \rangle \: \langle \hat{E}^\dagger(t+\tau) \:
    \hat{E}(t+\tau)\rangle}\\&=\frac{\Gtwo(t,t+\tau)}{\Gone(t,t)\Gone(t+\tau,t+\tau)},
  \end{align}
\end{subequations}
where $ \hat{E}$ are field amplitude operators. These quantities are,
respectively, field and intensity correlations at times $t$ and
$t+\tau$ at the same spatial location. Since for large time delays
realistic fields are uncorrelated, $\gtwo(\infty) = 1$. The behaviour
for finite delays depends however on the actual statistics of the
field. If $\gtwo$ increases around $\tau=0$, the field is
said to be bunched.

A chaotic field in an interval short with respect to its coherence
time $\cohtime$ will show thermal statistics, that is the distribution
$P_n$ of the number of photons is
\begin{equation}
  P_n = \frac{\nu^n}{ (\nu+1)^{(n+1)}},
\end{equation}
where $\nu$ is the average number of photons in the interval.  Chaotic light, for which $\gtwo(0)=2$, is therefore bunched. In contrast, the
output of an ideal mono-mode laser displays Poissonian behaviour, that
is $\gtwo(\tau) = 1$.  Hence, in a short time interval, the probability of
counting two photons for chaotic light is twice as large as that
expected for Poissonian statistics.

With the picture of an \spdc process as a spontaneous emission of pairs from
the splitting of pump photons, it could be
na\"ively believed that the statistics of pairs is similar to that of the pump
field\comment{, which is Poissonian when coherent pumping is used}. It is
however known \cite{Carusotto75} that linear interaction in a resonant medium
affects incident light in such a way that it becomes a mixture of the initial
\comment{(attenuated or amplified)} ``signal'' field and an additional
thermal field from spontaneous emission. If the incident field is absent (or
already thermal), the output field is therefore exactly thermal. In analogy,
since the \spdc process corresponds to optical amplification without input
signal (as both signal and idler fields are initially in their vacuum states),
the statistics of pairs needs not mimic that of the pump, and thermal
behaviour is to be expected.

As for incoherent light sources, the quantum interpretation of \spdc
bunching relies on constructive two-photon interference between
probability amplitudes for the paths of two indistinguishable photons
\cite{Fano61, Ou-book-07}. It has been shown that this process can
equivalently be interpreted as a pair emission stimulated by another
down-converted pair \cite[ch.7]{Ou-book-07}.

That the statistics of one arm of a mono-mode \spdc source is thermal
was theoretically first proven in 1987 by Yurke and Potasek
\cite{YurkePotasek87}, who derived from the pure state of the combined
single-mode down-converted fields the density operator of the mixed
state of one field only by tracing out either signal or idler
field. In this model, the quantum state $\ket{\phi}$ of the combined
signal-idler fields in the Fock state expansion reads
\begin{multline} \label{eq:evol_vacuum_monomode}
  e^{i (\eta\, a^\dagger b^\dagger + \eta^*\! a b)} \vacuum
  = \text{sech}|\eta| \sum_{n=0}^{\infty}
  \left(\frac{\eta}{|\eta|}\text{tanh}|\eta|\right)^n \! \ket{n}_{si} \\
  = (1-\tfrac{1}{2}|\eta|^2)\vacuum
  + \eta\ket{1}_{si} + \eta^2 \ket{2}_{si} + O(|\eta|^3),
\end{multline}
where $\eta$ is a parameter proportional to the pump amplitude,
$a^\dagger$ ($b^\dagger$) is the creation operator for the idler
(signal) mode with fixed polarisation, and $\ket{n}_{si} = \frac{1}{n!}
(a^\dagger b^\dagger)^n \vacuum$ is a Fock state with $n$ pairs.  By
tracing over one field, say $i$, the dominant signal-idler correlation
is suppressed, and a mixed density matrix is obtained for the residual
field that displays exactly thermal statistics:
\begin{equation} \label{eq:thermadens}
  \rho_{s}= \sum_{n=0}^{\infty} P_n \ketbra[s]{n}{n} =
  \sum_{n=0}^{\infty}\frac{\nu^n}{(\nu+1)^{(n+1)}} \ketbra[s]{n}{n},
\end{equation}
where $\nu = \text{sinh}^2|\eta|$ and $P_n$ is the probability of
finding exactly $n$ signal photons. The \spdc process is
  nevertheless inherently quantum mechanical, and its ``thermality''
  is not an effect of the trace operation, but originates directly
  from the nature of the pair production process
  \cite{YurkePotasek87}.

Since \spdc production is far from being mono-mode, Yurke and
Potasek's analysis was not realistic; however, the thermal nature of a
multi-mode \spdc process was later confirmed by Tapster and Rarity
\cite{TapsterRarity98}, and in a more general way by Ou, Rhee and Wang
\cite{Ou97, ORW99}, who analysed the case of sources with {\em pulsed
  pumping}.  In their derivation, what is actually measured is the
pulse integral of $\Gone$ and $\Gtwo$, {\em i.e.}, the probabilities
of one or two photons in the pulse. With time resolution limited to
the whole pulse, one must resort to test the ``bunching excess'',
which was shown \cite{ORW99} to be
essentially the ratio of the coherence time $\cohtime$ of the photons
in the inspected arm and the duration $\pumptime$ of the pulse, when
$\cohtime \ll \pumptime$. Shorter pump pulses then increase the
excess, but, contrary to intuition, the limit of a very short pulse is
not sufficient alone to guarantee a full bunching peak ($\gtwo(0)=2$). In
\cite[ch.7]{Ou-book-07} it is shown that this condition can only be
achieved with the help of narrow spectral filtering.

SPDC sources have also been implemented with continuous pumping, using
{\em continuous wave} (cw) lasers, which are both more practical and
more affordable than femto-second (fs) pulsed SPDC sources. The field emitted by cw
sources, and its statistical properties, is still relatively
unexplored. Theoretically, it is usually assumed that resolving times
are so much larger than the coherence time of the down-converted field
that all relevant quantities are not accessible.  It has been stated
\cite{ACLMWS08,LTS95,SHMIK06,MSNI06} that the
statistics of \spdc as a multi-mode process, {\em i.e.} when the
pulse duration of the laser is much longer than the coherence time of
the produced photons, will lead to Poissonian statistics. Careful
experimental design and data analysis, as shown in this paper, reveal
however the inherent thermal statistics, hidden by an apparent
Poissonian behaviour.

\subsection{Current experimental evidence}
\label{sec:experiments}
\newcommand{\smalltitle}[1]{\underline{\em\smash{#1}}\,}

Photon statistics in \spdc processes can be directly accessed using a
Hanbury Brown-Twiss setup by looking at photon emission in a single
output arm only, {\em i.e.} by measuring the $\gtwo$ of the signal or
idler arm. In such an investigation however, jitter and arrival-time
discretisation smear the result. Due to the a\-forementioned
disproportion between the time resolution of available single-photon
detectors and the \spdc coherence time, most experimental
investigations have exploited fs-pulsed sources and strong filtering to
increase artificially the coherence time of the measured
field. Positive evidence in single-photon regime was first reported by
Tapster and Rarity \cite{TapsterRarity98}, who compared quantitative
predictions based on a specific experimental hardware with
measurements performed with various filters; with the narrowest one, a
very convincing bunching peak of $1.85$ could be
demonstrated. \comment{Their predictions rely however on unverified
  assumptions on pulse shapes and filter transmissions, which are
  potentially responsible for imperfect agreement between theory and
  experimental data.} Six years later, de Riedmatten \etal
\cite{RSMATZG04} (almost) replicated the experiment, with similar
shortcomings and results, concluding that the amount of bunching
depends only on the ratio $\cohtime / \pumptime$. In both cases an
apparent transition from thermal to Poissonian statistics was observed
when $\cohtime$ was varied below $\pumptime$. \comment{An analogue
  experiment was also reported by Tsujino \etal \cite{THTS04}.}
S\"oderholm \etal\cite{SHMIK06} \comment{same authors as for
  \cite{MSNI06}} studied the dependence of single- and double-click
rates on pump power in the two cases $\cohtime \sim 1.75\pumptime$ and
$\cohtime \sim 0.6\pumptime$, and found good agreement with a thermal
model in the first case, and an intermediate behaviour in the second
case. On the other hand, Mori \etal \cite{MSNI06} studied a source
with very long pump pulses ($40$ps) and broad filtering leading to a
rather short coherence time $\cohtime$ ($\sim 120$fs), and failed to
detect any significant bunching. Similarly, in a recent article,
Avenhaus \etal \cite{ACLMWS08}, using $60$ps laser pulses, observed
Poissonian statistics for the \spdc process ($\cohtime$ not stated).

In an experiment where, contrary to the single-photon regime,
several photons per pulse are created, Paleari \etal \cite{PAZB04}
were able to demonstrate, with careful fitting of the measured
distribution, a thermal behaviour, although their ratio
$\pumptime/\cohtime$ was rather unfavourable (in the range
$10$-$20$). Vasilyev \etal\cite{VCKA98} used a strongly filtered
optical homodyne tomography to extract the photon-number distribution
of a \spdc process and compare it with a thermal one, finding almost
perfect agreement\footnote{Homodyne detection with a $10$MHz bandpass
  filter, hence {\em extremely} tight filtering.}.

\spdc sources have also been implemented using {\em continuous-wave}
(cw) pumping; this alternative approach\comment{which is equally
  interesting from a technological point of view} is still relatively
unexplored due to the technical difficulty of measuring very short
correlation times. Larchuk \etal\cite{LTS95} tried to investigate
sources in the single-photon regime with a single detector and delay lines
but failed to obtain any evidence because the dead time, about
$1\mu$s, makes the apparatus blind to the interesting region $\tau
\lesssim 1$ps. Super-Poissonian behaviour was found by Zhang
\etal\cite{ZKW02} by directly observing the photocurrents of
orthogonally polarised twin beams from a continuously driven \ktp
crystal in a cavity. \comment{, but this experiment of course is very
  far from single-photon regime and, due to the extremely narrow
  observed bandwidth, it is essentially mono-mode.}

 Measurements of the marginal, {\em i.e.}  one-arm, \spdc field are
 not to be confused with measurements of the better known statistics
 of the whole bi-photon state, as well as conditioned statistics,
 which have been experimentally confirmed, among others, by
 \cite{HPHP05,WSDY06,ASW06,TengnerLjunggren07}.

\comment{Measurements of the marginal, {\em
  i.e.}  one-arm, \spdc statistics are not to be confused with those
of the statistics of the whole two-arm state. Since in most of those
experiments $\pumptime$ is significantly larger than $\cohtime$,
thermality is difficult to spot unless explicitly looked for, and
most analysis are happy with observing an approximate Poissonian
distribution, which is of little interest for the subject of this
article. This is the case for instance of Haderka \etal\cite{HPHP05},
Waks \etal\cite{WSDY06}, Achilles \etal\cite{ASW06}, Tengner and
Ljunggren (with a cw pump) \cite{TengnerLjunggren07}, and Avenhaus
\etal\cite{ACLMWS08}. Another type of experimental investigation, with
a cw setup very similar to ours, focuses on one-arm statistics {\em
  conditioned} on a detection in the other arm \cite{BCRLW09}, but
their results and analysis does not relate to \spdc one-arm bunching
and is therefore not relevant here. }


\subsection{Quantum key distribution (QKD) and \spdc bunching}
\label{sec:QKD}
Sources based on continuously pumped \spdc are often employed in
quantum cryptographic devices, generating the multi-photon pair state
of Eq.\ref{eq:evol_vacuum_monomode}. Correlations between multi-photon
pairs constitute a potential threat to the security of QKD as the exchange
is then susceptible to a photon-number-splitting attack
\cite{BLMS00}. The case of {\em entangled-photon} sources~\cite{TPH09} is still
relatively unexplored, and its solution depends on a thorough
understanding of the extent to which multiple \spdc pairs are
correlated.

In entangled-photon sources, the state, to first order,
corresponds to a pair completely entangled in the information-carrying
degree of freedom (polarisation); {\em e.g.} for type-II \spdc $\ket{\psi^{-}} = (\ket{H_s,V_i} -\ket{V_s,H_i})/\sqrt{2}$.
When the second order of the expansion (Eq.\ref{eq:evol_vacuum_monomode}) is considered (two pairs), two limiting cases emerge: {\em sister} pairs are in well distinguishable temporal
modes, and are therefore completely uncorrelated in polarisation, with
all polarisation combinations equally likely,
\begin{equation}
  \label{eq:sisters}
  \ket{\textrm{sisters}} =
  \frac{ \ket{2H,2V} + \ket{2V,2H} - \sqrt{2} \ket{HV,VH} }{2},
\end{equation}
while {\em twin} pairs are in the same temporal mode (within $\cohtime$), and
\begin{equation}
  \label{eq:twins}
  \ket{\textrm{twins}} =
  \frac{ \ket{2H,2V} + \ket{2V,2H} - \ket{HV,VH} }{\sqrt{3}}.
\end{equation}
For twin states it has been shown that the probability that the two
signal photons have the same polarisation is $\frac{2}{3}$
\cite{DusekBradler02}; therefore, whereas no kind of
photon-number-splitting attack is possible with sister states, twin
states are potentially dangerous for quantum key distribution. The
situation is analogous for type-I \spdc.

The works of Tsujino \etal\cite{THTS04} and Ou \cite{Ou05} have shown
that polarisation correlation and one-arm bunching are two sides of the same
phenomenon: the peculiar properties of the state space of identical particles
imply that, for delays shorter than $\cohtime$, signal photons with the same
polarisation occur more frequently than expected for classically uncorrelated
objects (as differently polarised photons).

Summarising, the study of one-arm $\gtwo$ in entangled-photon
sources \footnote{As opposed to studying $\gtwo$ in
  attenuated lasers and heralded-photon sources.}  gives information
about the maximum possible extent of the security threat posed to
quantum key distribution by higher orders of \spdc.



\section{Experimental setup and data acquisition}
\label{sec:setup}

\begin{figure}[t!]
  \begin{center}
    \includegraphics[width=\columnwidth]{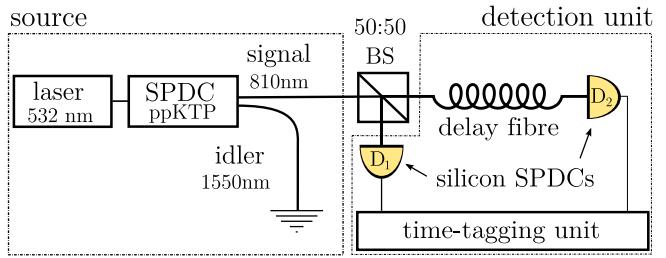}
  \end{center}
  \caption{\label{fig:setup} (colour online) The experimental setup
    consists of a \spdc source of photon pairs, a balanced beam splitter
    and two single-photon silicon detectors, combined in a Hanbury Brown
    and Twiss \cite{HBT56} configuration.}
\end{figure}

In our \spdc source \footnote{This polarisation-entangled source was
  designed for quantum cryptography, and will be the subject of a future
  publication [M.~Hentschel, H.~H\"{u}bel, A.~Poppe and A.~Zeilinger, {\em Three-color Sagnac source of polarization-entangled photon pairs}, 2009, arXiv, to be submitted to Optics Express]. Here the source was
  employed in unentangled mode, with down-converted fields having fixed
  linear polarisation.} (Fig.\ref{fig:setup}), a $532$nm cw laser pumps a
single $30$mm temperature-stabilised nonlinear crystal
(periodically-poled KTP) for \mbox{type-I} down-conversion (532nm $\rightarrow$ 810nm $+$ 1550nm).  The
signal and idler photons have fixed polarisations, central
wavelengths of respectively $810$nm and $1550$nm, and are emitted
collinearly. The signal coherence time, as measured with a Michelson
interferometer (Fig.\ref{fig:interferogram}), is $\tau_c \sim 2.8$ps
\fwhm, corresponding to a bandwidth of less than $1$nm.

Signal and idler photons are separated according to their frequency,
and coupled into single-mode optical fibres. Signal photons (at
$810$nm) are then recollimated and sent to a non-polarising balanced
beam splitter. The resulting beams are collected into multi-mode
fibres. The coupling efficiency is greater than $90\%$ as the light is
coupled from single-mode fibres into multi-mode ones. The multi-mode
fibres direct the beams into a {\em PerkinElmer} \textsc{\footnotesize
  SPCM-AQ4C} single-photon detector array. The parameters of each
detector are as follows: quantum efficiency at $810$nm $\sim 50$\%,
dark-count rate $< 500$Hz, dead time $\sim 50$ns, and detector
saturation level above $1$MHz. All detector clicks are recorded by a
time-tagging unit (TTU) (\emph{TT8, smart systems Division, ARC}) with
a common time basis and an intrinsic resolution of
$\tau_\textrm{{\tiny \textit{TT}}}= 82.2$ps. Time tags are then
processed to define coincidences.

One of the two fibres is a $100$m spool, so that it operates as a delay
line of approximately $500$ns (this delay is removed by software during
data analysis). The purpose of the delay line is to make photons that
happen to come with a picosecond delay to impinge on detectors at
sufficiently different times to avoid electronic cross-talk and increase
sensitivity.

The timing jitter of the overall detection unit was characterised by
measuring correlations in arrival times of signal-idler photon-pairs
from an additional degenerate \spdc source at $810$nm (405nm
$\rightarrow$ 810nm $+$ 810nm). This auxiliary photon-pair source is
based on a Sagnac configuration \cite{Fedrizzi07}. For the present
measurement of jitter, it was operated to yield $\sim 1$MHz detected
single rates, below detector saturation.  The coincidence rate of the
photon-pair source was approximately $100$kHz. The combined jitter of
the detection unit is estimated to be $\tau_j =640 \mbox{ps} \gg
\tau_c$. This estimated FWHM is however not used in further
experimental analysis, and no assumption on the shape of the jitter is
needed. Instead, the fully normalised jitter curve, as seen in the
cross-correlation in Fig.\ref{fig:jitter}, is used to characterise the
detection unit, the only extracted parameter being the value of the
area under the jitter curve ($611$ps). Several measurements of the
jitter histogram were taken, and the deviation on the jitter area is
estimated to be less than $3\%$.

A cross-correlation histogram is obtained from re\-cord\-ed detector
clicks. A coincidence from detector $D_1$ to detector $D_2$ with delay
$\Delta t = t_2 - t_1$ is counted if $D_1$ clicked at time $t_1$, and
$D_2$ clicked at time $t_2 > t_1$, irrespective of any other click.
Coincidences with the role of detectors reversed are defined
accordingly, and the two functions are joined by arbitrarily defining
the delays $\Delta t$ from $D_2$ to $D_1$ as negative.

The coincidence density for uncorrelated photons, {\em e.g.} when
$\Delta t$ is large with respect to any coherence time, is the product
of the individual experimental detection rates $\lambda_1$ and
$\lambda_2$ of $D_1$ and $D_2$ (which include the effects of dead
times and dark counts). Therefore, every bin of the coincidence
histogram will contain, on average, and in the case of uncorrelated
photons, $C = \lambda_1 \lambda_2 \,\tau_\textrm{{\tiny
    \textit{TT}}}\, T$ counts, where $T$ is the duration of a
data-taking run. \comment{A correction is provided for the case of
  non-constant source intensity.} Since $C$ contains both single rates
$\lambda_1$ and $\lambda_2$, it is directly proportional to the square
of the pump power.


\section{Model of expected photon bunching and measurements}
\label{sec:model}

Since the jitter $\tau_j$ of the detection unit is much larger than the
coherence time $\tau_c$ of the down-converted light, any $\gtwo$ peak will
be strongly smeared. A rough estimate for the residual peak height is given
by the ratio of the coherence time to the jitter $\tau_c / \tau_j \sim 4
\times 10^{-3}$.

For a more accurate calculation, we model how a theoretical thermal
bunching peak ($\gtwo(0) = 2$) will be transformed due to our
instruments and then compare it with the experimental result.
\spdc-bunching is characterised by the same relation between $\gtwo$ and the
first order coherence $\gone$ that holds for chaotic light
\cite{Loudon_book}, namely:
\begin{equation} \label{eq:g1g2}
  \gtwo(\tau)=1+|\gone(\tau)|^2
\end{equation}


\begin{figure}[t!]
  \includegraphics[width=\columnwidth]{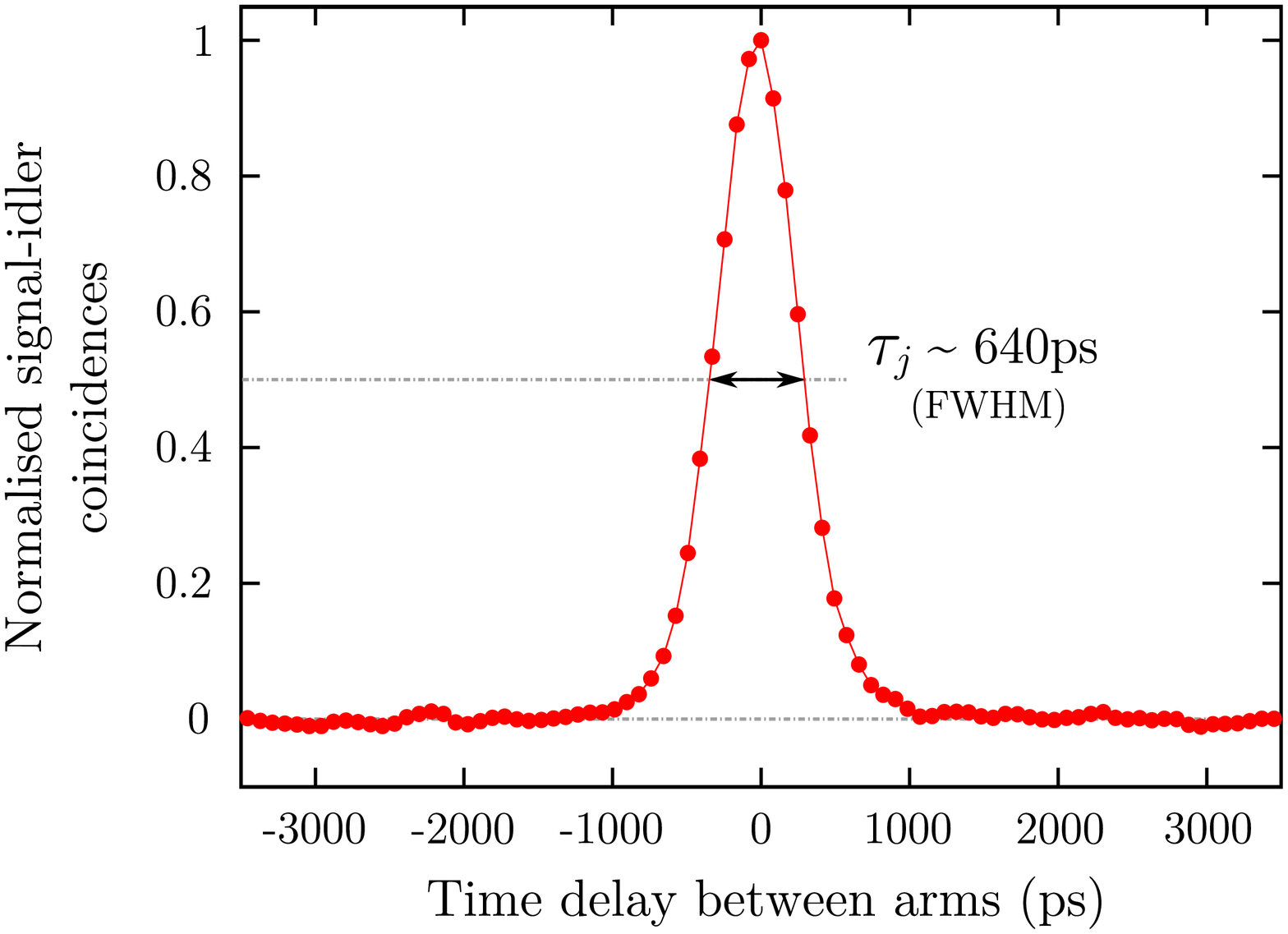}
  \caption{\label{fig:jitter} (colour online) Second-order correlation
    function between the idler and signal arms of an \textit{auxiliary}
    degenerate \spdc source at $810$nm (405nm $\rightarrow$ 810nm $+$ 810nm), rescaled to unity. \comment{This
      $\gtwo(0)$ between the signal-idler photons is very large because the
      \spdc field is essentially a collection of pairs.} Since the time
    correlation between photons of the same pair is tight
    (sub-picoseconds), the \fwhm, approximately $640$ps, is essentially an
    estimate of the combined jitter of the two silicon detectors and the
    time-tagging unit. $\;\;\;\;\;\;\;\;\;\;\;\;\;\;\;$ }
\end{figure}

\begin{figure}[t!]%
  \includegraphics[width=0.96\columnwidth]{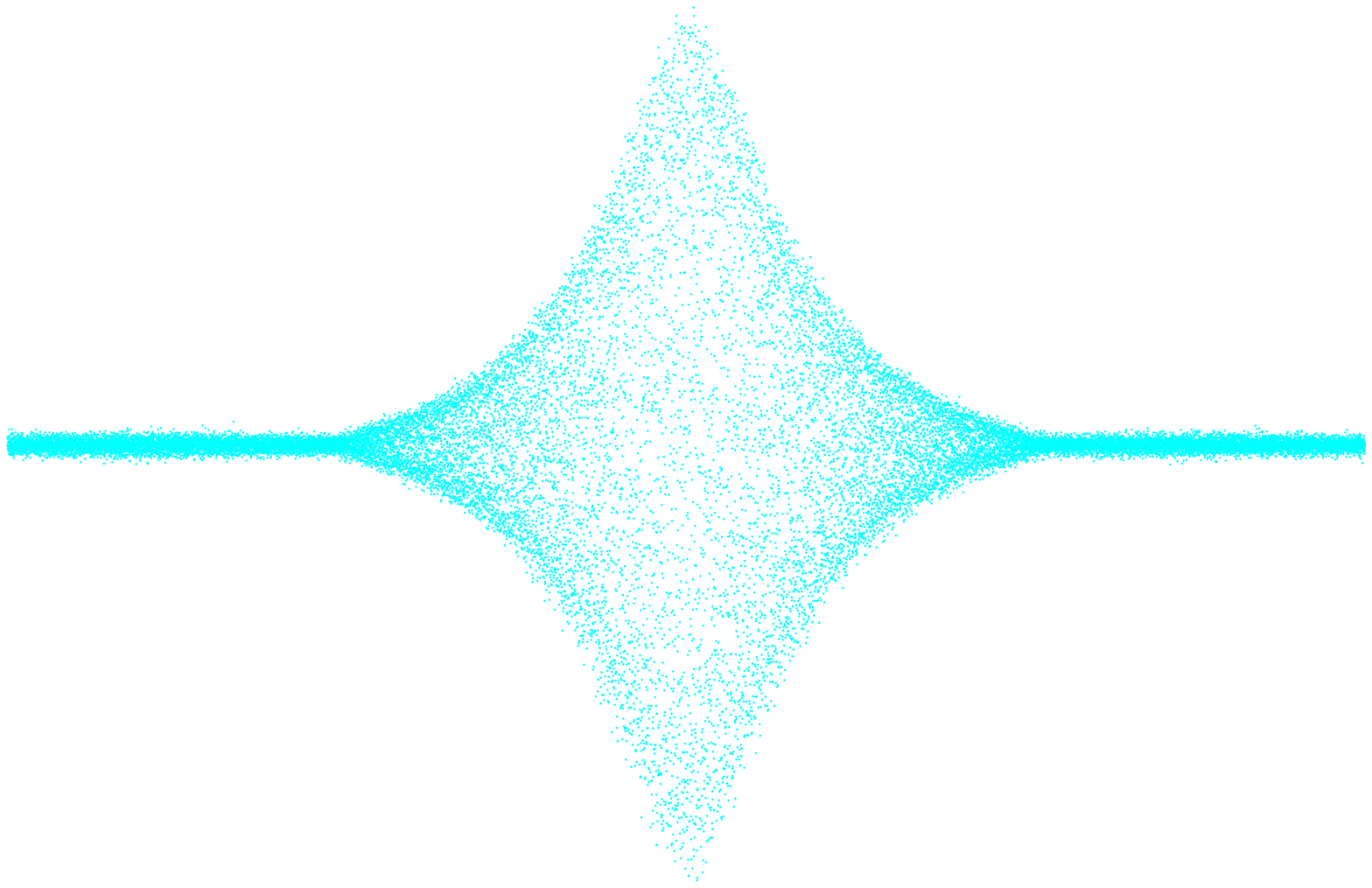}%
  \llap{\includegraphics[width=0.96\columnwidth]{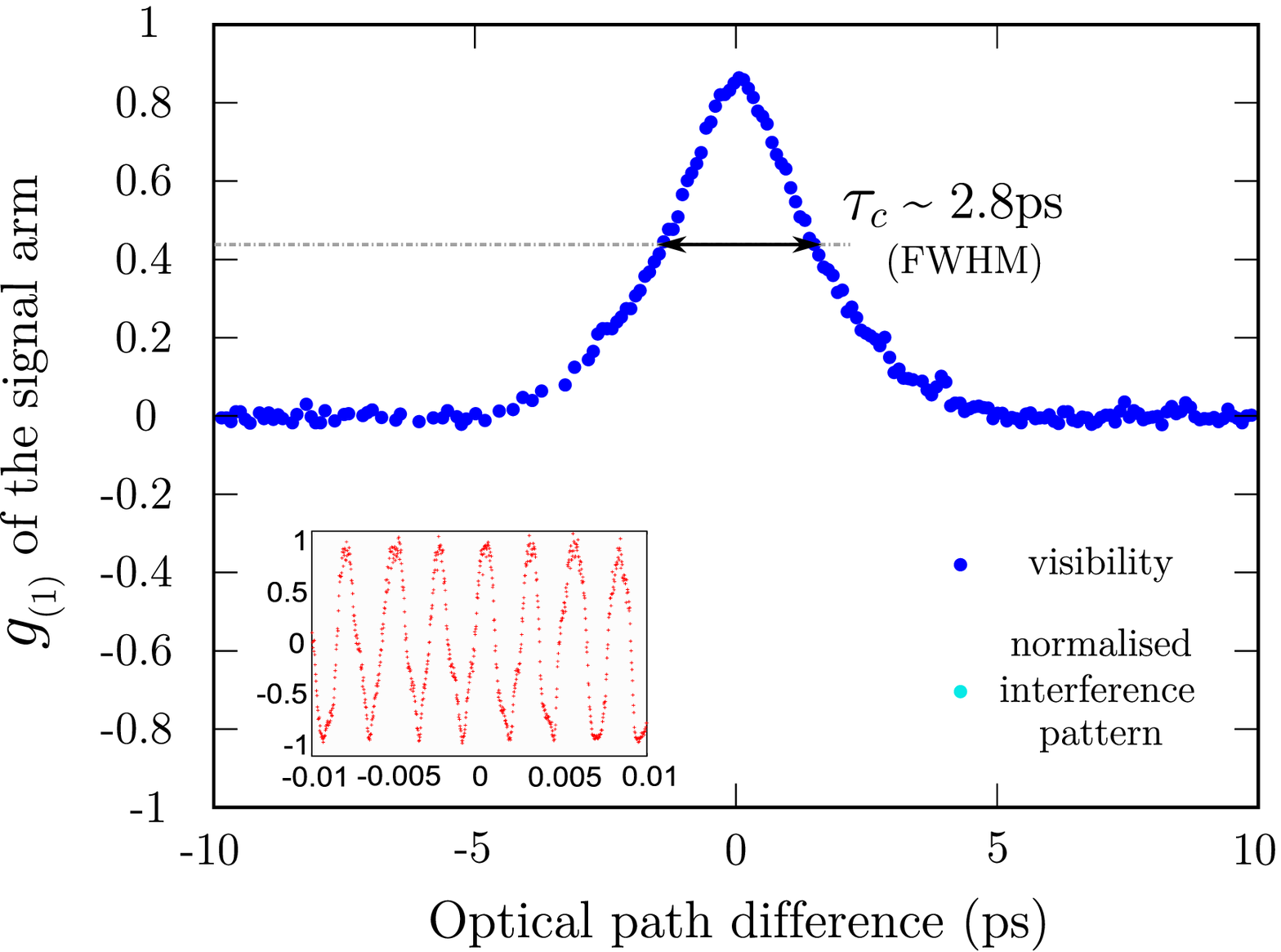}}
  \caption{\label{fig:interferogram} (colour online) $\gone$ of the
    $810$nm arm of the \spdc source as measured in a Michelson
    interferometer vs.  optical path difference. $\gone$ was
    calculated from the fringe-visibility of the interferogram. The
    experimental visibility at zero optical path difference is
    slightly less than 1.  When the maximum of the visibility is
    rescaled to one, the total area under the {\em squared} envelope
    is $\sim 2.2$ps. The inset shows the visibility fringes around the
    central part of the interferogram.}
\end{figure}

The $\gone$ of the signal beam was measured in a Michelson
interferometer and can be seen in Fig.\ref{fig:interferogram}. The
setup for this measurement consisted of a Michelson interferometer
with a motorised movable mirror and some limiting apertures in both
arms to increase visibilities. The maximum visibility of the
interferogram was measured to be $90 \%$. The mirror is set to move at
a constant speed of $0.002$ mm/s over a range of $3$mm, and detection
counts were recorded over integration windows of $2$ ms. At unequal
path lengths, the detector count rate at the output of the
interferometer was in the range of $200$kHz. The interference fringes
in Fig.\ref{fig:interferogram} were obtained from the raw data by
subtracting the background rates (around $5$kHz), and are displayed
with an averaging of 30 data points to reduce fluctuations away from
the centre of the interferogram.

The envelope of the interferogram as shown in
Fig.\ref{fig:interferogram}, that is $|\gone|$, is calculated from the
visibility of the interference fringes in the raw data (see inset in
Fig.\ref{fig:interferogram}). An estimate of the coherence time
$\tau_c \sim 2.8$ps is obtained from the graph. After rescaling the
data points of the $|\gone|$ envelope to compensate for the reduced
visibility ($90\%$) in the interferogram, an area of $2.17$ps was
found for the {\em squared} envelope $|\gone|^2$. The error on this
value can be estimated by considering the area of several interference
scans in identical experimental conditions; a conservative estimate is
a $4\%$ error on the calculated area for the squared envelope.

To model the washed out $\gtwo$ peak, we note that the total number of
coincidences in the excess peak is preserved despite the jitter. Hence
we expect to observe a smeared peak with the same temporal profile as
that in Fig.\ref{fig:jitter}, but with an area equal to that of
$|\gone|^2$ \footnote{This analysis does not take into account spatial
  (transversal) coherence. Due to the use of single mode fibres, the
  setup was sufficiently optimised for such considerations to have no
  major implications in the model.}. This method does not rely on any
\emph{a priori} knowledge of the actual shapes of $\gone$ or $\gtwo$,
but only on a ratio of areas. The estimated coherence time $\tau_c$
and jitter $\tau_j$ of the detection unit from the FWHM of their
respective histogram are not used in the calculation, neither is any
assumption on curves shapes necessary.

The correlation function $\gtwo$ of the field from one arm of a \spdc
source was measured using the Hnbury Brown-Twiss-like setup introduced in the
previous section. A cross-correlation histogram for positive and
negative delays was obtained from detector clicks and the bunching
peak emerged in the region $\Delta t \sim 0$ over the plateau of
accidental coincidences. The average count number of the plateau was
then used to normalise the histogram.  Experimental data were
collected during a $16$ hours data-taking session with an average
count rate of $\sim 1$ MHz in each detector (below saturation), giving
a plateau count of $N \simeq 4.4\times10^{6}$ events per bin.  The
observed rates used for jitter characterisation were set to be
identical to the rates in this measurement, since detector jitter is
dependent on actual count rates. The double pair term in
Eq.\ref{eq:evol_vacuum_monomode} scales with the square of the pump
power, as does the plateau (see the estimate $C$ at the end of
Sec.\ref{sec:setup}), hence the normalised $\gtwo$ function is
independent of pump power. The same argument can be used to prove the
peak to be independent of optical losses in the setup.

\begin{figure}[t!]
  \includegraphics[width=\columnwidth]{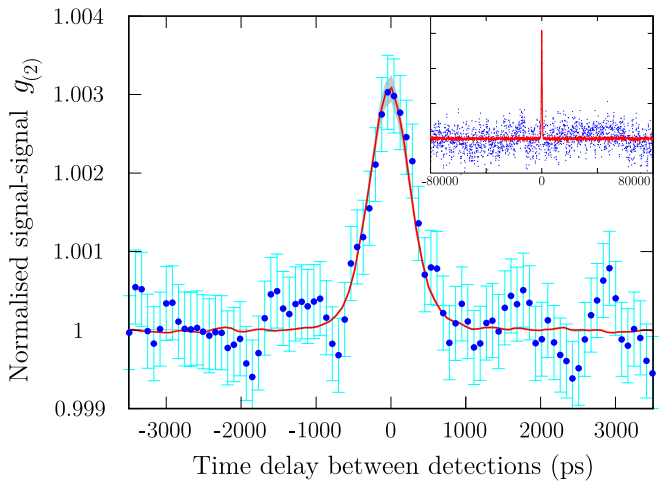}
  \caption{\label{fig:g2peak} (colour online) Second-order correlation
    function (signal-signal $\gtwo$) of the $810$nm arm of a \spdc
    source, normalised over a measured plateau of $N$ counts in a
    bin. The solid line represents the washed-out bunching peak
    (original height of 2), obtained from the jitter plot rescaled to
    an area of $2.17$ps. Error bars, representing statistical
    fluctuations of the measured data, are set to $\sqrt{N}$. The
    shaded grey area, only visible at the very top of the peak, represents the uncertainty in the model
    prediction. The inset shows an enlarged section of the plateau.}
\end{figure}

In Fig. \ref{fig:g2peak}, we show a comparison of the measured data to
the expected $\gtwo$ of our model. Data points correspond to the
experimental $\gtwo$ histogram of the marginal \spdc field. The solid
line is the expectation for the bunching peak of a field with thermal
statistics. This line was obtained by vertically rescaling the jitter
function shown in Fig.\ref{fig:jitter} to yield the same area as
$|\gone|^2$. The comparison does not involve any free parameter to be
fitted: the theoretical model is only based on the specific and
independently measured parameters of the setup.

 The peak was shifted laterally by $\sim 50$ps for better comparison
 with the experimental data; this is however less than the intrinsic
 resolution $\tau_\textrm{{\tiny \textit{TT}}}$. Statistical fluctuations of the number of
 events in a bin of the histogram are of the order of $\sqrt{N}$. The
 cumulative error of $\pm 5\%$ on the expected $\gtwo$ curve, arising
 from a conservative error estimation from the jitter and $|\gone|^2$
 data, is shown in Fig.\ref{fig:g2peak} as a shaded area. Even this $10\%$
 error margin fits the data points, and therefore the good match of
 the predicted peak is not accidental.

The emerging bunching from the \spdc field is in excellent agreement with
the theoretical estimation of a thermal statistics. The peak protrudes 6
standard deviations above the Poissonian limit of $\gtwo = 1$ and is
therefore experimentally confirmed.

\section{Conclusions}
\label{sec:conclusions}

We experimentally measured the marginal temporal second-order
correlation function $\gtwo(\tau)$ of \spdc light with
continuous wave excitation in the single-photon regime. To our
knowledge, the results represent the first direct observation of
photon bunching for such a field, and strongly confirm its thermal
character. Even with an experimental excess peak $\gtwo(0)$ as low as
$10^{-3}$, our experimental accuracy is sufficient to properly observe
the residual bunching peak, which can be modelled solely on the
experimental timing jitter and the output of a first-order
interferometric measurement ($\gone$). With much higher timing
precision (e.g. lower detector jitter), it is expected that a full
peak with the shape given by Eq.(\ref{eq:g1g2}) is recovered. The
relevant parameter is the ratio between the coherence length $\tau_c$
and the measurement resolution (the jitter $\tau_j$ in the cw
case). \comment{If a timing precision of $\tau_j \lesssim \tau_c$ is
  not achieved, the $\gtwo$ peak will be washed out and possibly
  drowned in statistical fluctuations.}

We therefore conclude that thermal photon statistics of the marginal
\spdc field can be observed both in the fs-pulsed and in the cw regime,
the latter being more practical and more affordable. A multimode
excitation does not change the statistics of \spdc but alters only the
experimental conditions, so that without a careful analysis an {\em
  apparent} Poissonian character is observed (as presented in some
works reviewed in section \ref{sec:experiments}), instead of the
inherent thermal statistics. We have shown in this paper that even
with far-from-ideal detection modules, and without any additional
spectral filtering, we could demonstrate as a proof of principle that
the true nature of the field remains accessible.

It is hoped that the results presented here stimulate further
investigations into the nature of \spdc light and its thermal
properties. For cw pumping, there is no clear consensus on the actual
state of multi-photon pairs, nor is its extension to entangled states
produced by cw pumping well understood. Similar experiments would
provide experimental data to confirm that bunching
implies polarisation correlations. In addition, cw pumping allows a
continuous temporal investigation in a time domain not accessible to
experiments with pulsed pumping and hence, with an improved timing
resolution, access to the shape of $\gtwo$, and not only to its integral
(a single data point) as in pulsed setups.

We would like to thank Michael Hentschel, Thomas Lor\"unser and Edwin
Querasser for technical support with the experiment, as well as
Momtchil Peev, Miloslav Du\v{s}ek, Thomas Jennewein and Anton
Zeilinger for motivation and advice. We also acknowledge fruitful
discussions with Saverio Pascazio and Bahaa Saleh. We acknowledge
financial support from the Austrian FWF (SFB15, TRP-L135).

\hbadness=10000
\bibliography{Doppelpaaren}

\end{document}